\documentclass[sigplan,screen, nonacm]{acmart}

\usepackage{listings}

\definecolor{codebg}{gray}{0.95}
\definecolor{codekw}{rgb}{0.10,0.25,0.55}
\definecolor{codecomment}{rgb}{0.30,0.45,0.30}
\definecolor{codestring}{rgb}{0.55,0.25,0.15}

\definecolor{codebg}{gray}{0.95}
\definecolor{llvmkw}{rgb}{0.10,0.25,0.55}   % structure: define, call, ret
\definecolor{llvmty}{rgb}{0.00,0.42,0.42}   % types: i64, i1, ptr
\definecolor{llvmop}{rgb}{0.55,0.15,0.35}   % opcodes: sub, and, icmp
\definecolor{llvmfn}{rgb}{0.65,0.30,0.05}   % @globals
\definecolor{llvmvar}{rgb}{0.35,0.25,0.60}  % %locals
\definecolor{cmtgray}{gray}{0.45}

\lstdefinelanguage{LLVM}{
  sensitive=true,
  alsoletter={\%@\_},
  morecomment=[l]{;},
  morekeywords={define,declare,musttail,call,ret,cc,void,entry},
  morekeywords=[2]{i1,i8,i16,i32,i64,ptr},
  morekeywords=[3]{sub,add,and,or,xor,icmp,select,eq,ne,
    load,store,br,phi,switch},
}

\lstdefinelanguage{AArch64}{
  sensitive=true,
  alsoletter={\_},
  morecomment=[l]{;},
  morekeywords=[3]{mov,and,add,sub,str,ldr,stp,ldp,b,br,bl,ret},
}

\usepackage{algorithm}
\usepackage{algpseudocode}
\usepackage{tikz}
\usetikzlibrary{arrows.meta,positioning}
\usetikzlibrary{positioning, fit, backgrounds, arrows.meta, shapes.geometric, backgrounds}
\usepackage{amsmath}

\usepackage{listings}
\usepackage{xcolor}
\usepackage{subcaption}
\usepackage{adjustbox}   % in preamble (TAPS-approved)

\usepackage[utf8]{inputenc}
\usepackage[LGR,T1]{fontenc}
\usepackage[greek.polutoniko,english]{babel}
\usepackage[scaled=0.82]{beramono}

\newcommand{\coolname}{\textsc{ElevatorII}}
\newcommand{\reg}[1]{\texttt{#1}}
\begin{document}

\title{\coolname{}: High-Performance Fully-Static Cross-ISA Binary Translation}


\author{Hongyu Chen}
\email{hongyc4@uci.edu}
\orcid{0009-0008-0834-8251}
\affiliation{%
  \institution{University of California, Irvine}
  \city{Irvine}
  \state{CA}
  \country{USA}
}

\author{James McGowan}
\email{jwmcgowa@uci.edu}
\orcid{0009-0007-2523-454X}
\affiliation{%
  \institution{University of California, Irvine}
  \city{Irvine}
  \state{CA}
  \country{USA}
}

\author{Michael Franz}
\email{franz@uci.edu}
\orcid{0000-0001-5911-2275}
\affiliation{%
  \institution{University of California, Irvine}
  \city{Irvine}
  \state{CA}
  \country{USA}
}


\begin{abstract}

The established method for running software compiled for one instruction-set architecture (ISA) on a different ISA is by emulation in conjunction with just-in-time (JIT) compilation. This requires run-time support (at least a JIT) on the target that needs to be trusted and whose compilation decisions are hard to predict or reconstruct.

We present \coolname{}, a fully static, ahead-of-time cross-ISA binary-to-binary compiler from x86 to AArch64 that challenges the established consensus. Our approach eliminates the need for run-time support on the target, removes the hard to reproduce effects of dynamic compilation, and provides the capability to exhaustively test the actual binary that is eventually executed. Our system is therefore less risky to deploy than emulators that have embedded JITs.% As we show in this paper, our technique is also competitive.

Our approach combines superset disassembly with observation-guided path selection. Superset disassembly overcomes the code vs. data differentiation problem inherent in disassembly by creating a separate control flow path for every possible interpretation of every byte in the original input binary. It guarantees that there is pre-translated code for every possible jump destination in the original text segment, even for computed jumps that are not statically decidable. Static and observation-guided path selection then creates additional optimized ``confirmed code paths'' that overlay the regular grid of control flow paths derived through superset disassembly.% We also separate hot and cold code for i-cache optimization.

Our prototype handles a wide range of representative workloads. On the 11 SPEC CPU 2026 Integer benchmarks that do not use C++ exceptions, \coolname{} runs at $2.5\times$ native AArch64 time, on par with Box64, the fastest dynamic translator, and $5\times$ faster than QEMU. It also translates binaries obfuscated with virtualization and branch functions, which no static disassembler can follow. Our approach trades execution performance for space: our output code is $51\times$ the size of code compiled from source. We alleviate this by separating cold code to a \texttt{.text.unlikely} ELF section: only 2-3\% hot code winds up in the main ELF section, while the rest of the code will likely never be loaded into memory.
%moving the rest into a much larger  ELF section.

%It comes with the benefit that translated code can be tested before shipping, needs no OS or runtime support, and is therefore less risky to deploy.

% Our prototype is able to handle a wide range of representative workloads. It does not yet handle C++ exceptions or multi-threaded binaries, but for the 14 of the 17 SPECrate 2026 Integer benchmarks that don't raise exceptions, our system outperforms the two leading, mature, state of the art dynamic translators, Box64 and QEMU. Our results show that trading space for speed is indeed an option in X-ISA BT, with the additional benefit that in our approach, the actual code that is eventually executed can be exhaustively tested before deployment, doesn't depend on any hard to reproduce dynamic behavior, doesn't require any OS or other runtime support, and is therefore less risky to deploy.

\end{abstract}

\maketitle

\section{Introduction}

% A substantial part of the world's infrastructure runs on legacy binaries---venerable compiled programs that are hard to maintain properly. Some of these legacy binaries can no longer be maintained at all because  part of the build pipeline that was used during their original creation is no longer operable, or because some part of the original source code has gone missing. The long term plan for such legacy binaries typically involves eventual replacement by rewriting the entire software from scratch, but doing so requires money and time.

% A substantial part of the world's infrastructure runs on legacy binaries, compiled programs whose source code or build pipeline is no longer available. When the processor they were built for goes out of production or must be replaced, rewriting them takes years, and something must run the existing binary on the new processor in the meantime. Dynamic binary translation does this by decoding and compiling the program while it runs. The cost is that the translator becomes part of the deployment. It must be installed on every target machine, the code that executes is generated at run time and differs from run to run, and a failure in the field can be hard to reproduce.

A substantial part of the world's infrastructure runs on legacy binaries, compiled programs whose source code or build pipeline is no longer available. The binary itself is what has been tested and trusted, and any rewrite, no matter how it is produced, is a new program whose behavior must be established from scratch. This becomes a problem when such legacy binaries need to be replaced at short notice, for example because a low-level software vulnerability has been discovered that cannot be mitigated easily, or because there is a supply shortage of the processor that the legacy binary requires for execution. In such cases, coming up with an effective temporary "stop-gap" solution to be able to continue using the existing binary might be very valuable, even if it comes at a certain cost, for example reduced execution performance. For some combinations of source and target ISA, the overheads introduced by binary translation might even be partially compensated by the availability of more modern implementations of the target ISA.

This paper explores using cross-instruction-set-architecture binary translation (X-ISA BT) as such a ``drop-in'' %, stop-gap''
solution to quickly adapt an existing legacy binary to a new target ISA. In particular, we focus on fully static translation which takes an existing \emph{input binary} for one processor (x86-64 in our case) and generates a functionally identical \emph{output binary} for a different ISA (AArch64 in our case). Fully static translation means that the output binary is a complete standalone replacement of the input binary  that doesn't require any additional translation support or fall-back mechanisms. In particular, the output binary can be tested exhaustively before it is shipped. This makes the fully-static approach less risky than dynamic or hybrid binary translation approaches in which the exact composition of the output program depends on run-time factors that are hard to predict in advance and hard to reproduce if something goes wrong. It also removes the requirement for the translation system to be trusted, since the actual output code that will eventually execute on the target can now be tested directly.

% Our prototype system, \coolname{}, scales to a wide range of representative workloads and demonstrates that fully-static cross-ISA binary translation is not only viable, but also surprisingly competitive. The two major constructs we currently don't support are C++ exceptions and multi-threaded binaries. 
% %The one major construct that we currently don't support is C++ exceptions. 
% Apart from these limitations, \coolname{} can take almost any dynamically linked x86-64 ELF binary and produce a corresponding AArch64 output binary. In our experiments, these output binaries run on par with Box64, the leading dynamic binary translator, and consistently outperform QEMU, the long-established general-purpose translator (Section~\ref{sec:runtime}).
% %the two leading dynamic binary translators Box64 and QUEMU, on all 14 of the 17 SPECint2026 benchmarks that don't use C++ exceptions.

Our prototype system, \coolname{}
scales to a wide range of representative workloads and demonstrates that fully-static cross-ISA binary translation is not only viable, but also  competitive. \coolname{} can take almost any single-threaded, dynamically linked x86-64 ELF binary and produce a corresponding AArch64 output binary. In our experiments, these output binaries run on par with Box64, the leading dynamic binary translator, and consistently outperform QEMU, the long-established general-purpose translator.% (Section~\ref{sec:runtime}).

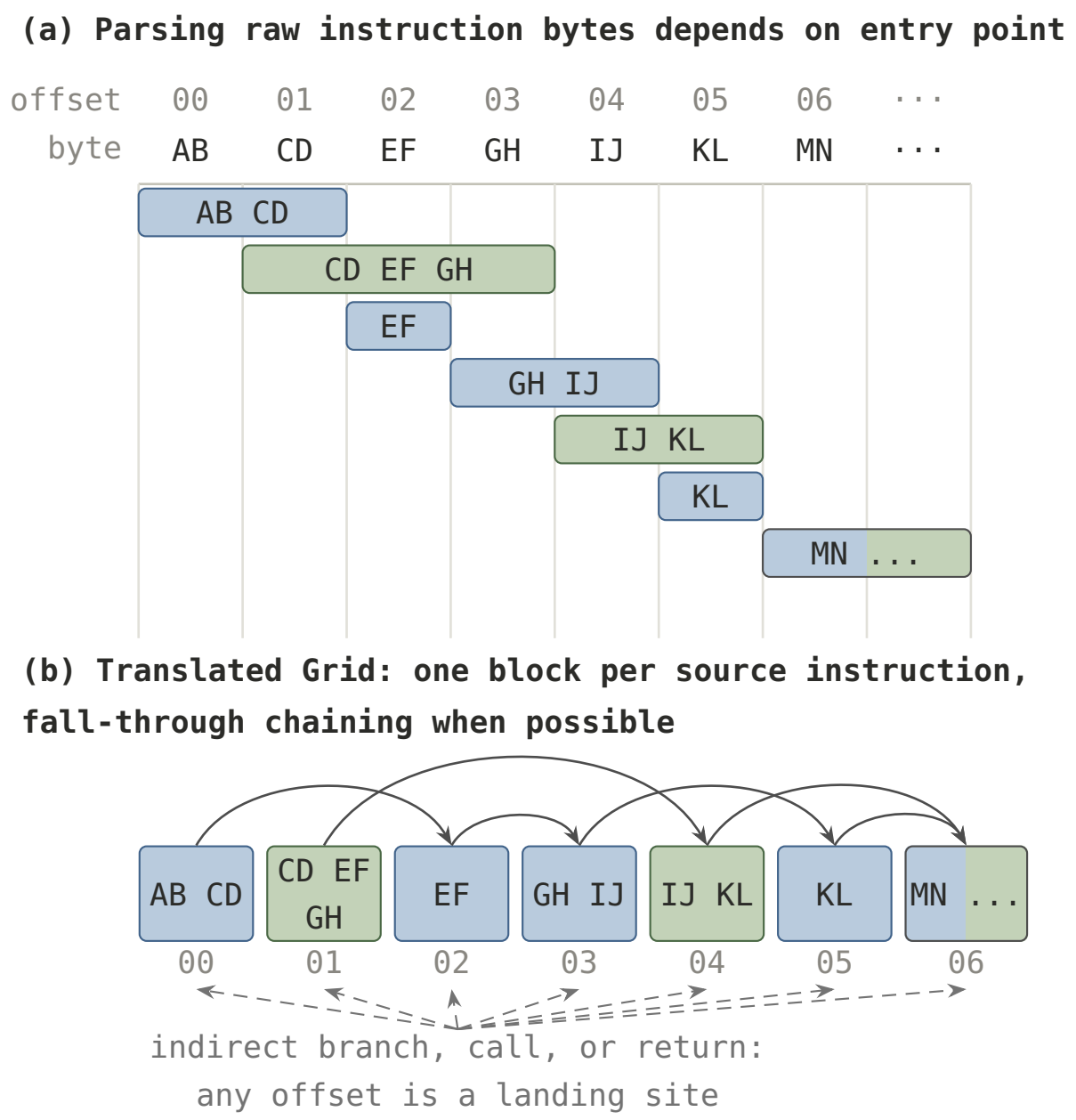
\begin{figure}[t]
  \centering
  \adjustbox{max width=\columnwidth}{% Superset disassembly and the grid, schematic, two panels.
% (a) Decoding starts at every byte offset, so the decodes overlap: the
%     decode at 00 consumes AB CD, the one at 01 consumes CD EF GH, ...
% (b) The grid: one block per byte offset, in address order. A block's
%     fall-through edge goes to the block at offset + length. Any address a
%     control transfer can reach has a block, so every indirect target lands
%     in the grid.
% Palette shared with figures/figstyle.py and slow-fast-paths-figure.tex.
% Fonts are \footnotesize (8pt) to meet the ASPLOS figure-text minimum;
% natural width ~215pt, so \adjustbox does not scale it down.
\definecolor{inkPrimary}{HTML}{2b2b28}
\definecolor{inkMuted}{HTML}{898781}
\definecolor{gridLine}{HTML}{e1e0d9}
\definecolor{baseLine}{HTML}{c3c2b7}
% One color per fall-through stream: the stream from 00 (AB CD, EF, GH IJ, KL)
% and the stream from 01 (CD EF GH, IJ KL). MN is reached by both.
\definecolor{pAfill}{HTML}{b9cbdd}\definecolor{pAedge}{HTML}{41638a}
\definecolor{pBfill}{HTML}{c3d2b8}\definecolor{pBedge}{HTML}{4a6845}
\begin{tikzpicture}[
  hdr/.style={font=\ttfamily\footnotesize, text=inkMuted, inner sep=0},
  byte/.style={font=\ttfamily\footnotesize, text=inkPrimary, inner sep=0},
  % (a) a decode of the input
  span/.style={rounded corners=1.5pt, line width=0.4pt},
  pA/.style={draw=pAedge, fill=pAfill},
  pB/.style={draw=pBedge, fill=pBfill},
  spantext/.style={font=\ttfamily\footnotesize, text=inkPrimary, inner sep=0},
  % (b) a translated block: two text lines so "CD EF GH" fits the 27pt pitch
  cell/.style={rounded corners=1.5pt, line width=0.4pt,
    font=\ttfamily\footnotesize, text=inkPrimary, align=center,
    minimum width=24pt, minimum height=20pt, inner sep=1pt, anchor=west},
  arr/.style={-{Stealth[length=4pt]}, black!70, line width=0.5pt},
  darr/.style={-{Stealth[length=4pt]}, densely dashed, black!55, thin},
  lbl/.style={font=\ttfamily\footnotesize, text=black!55, align=center, inner sep=1pt},
  panel/.style={font=\ttfamily\footnotesize\bfseries, text=inkPrimary, anchor=west, inner sep=0},
]
% ---------------- (a) overlapping decodes ----------------
\begin{scope}[x=22pt, y=-12pt]
\node[panel] at (-1.15, -1.2) {(a) Parsing raw instruction bytes depends on entry point};
\node[hdr, anchor=east] at (-0.15, 0)   {offset};
\node[hdr, anchor=east] at (-0.15, 0.9) {byte};
\foreach \i/\a/\b in {0/00/AB, 1/01/CD, 2/02/EF, 3/03/GH, 4/04/IJ, 5/05/KL, 6/06/MN} {
  \node[hdr]  at (\i+0.5, 0)   {\a};
  \node[byte] at (\i+0.5, 0.9) {\b};
}
\node[hdr]  at (7.5, 0)   {\ldots};
\node[byte] at (7.5, 0.9) {\ldots};
\draw[baseLine, line width=0.5pt] (0, 1.5) -- (8, 1.5);
\foreach \i in {0,...,8} { \draw[gridLine, line width=0.5pt] (\i, 1.5) -- (\i, 9.5); }
% decode at offset #1 consumes #2 bytes: a box spanning those byte columns
\newcommand{\dec}[4]{%
  \draw[span, #4] (#1, 2+#1-0.42) rectangle (#1+#2, 2+#1+0.42);
  \node[spantext] at (#1+#2/2, 2+#1) {#3};}
\dec{0}{2}{AB CD}{pA}
\dec{1}{3}{CD EF GH}{pB}
\dec{2}{1}{EF}{pA}
\dec{3}{2}{GH IJ}{pA}
\dec{4}{2}{IJ KL}{pB}
\dec{5}{1}{KL}{pA}
% MN is reached by both streams: half of each color
\begin{scope}
  \clip[rounded corners=1.5pt] (6, 8-0.42) rectangle (8, 8+0.42);
  \fill[pAfill] (6, 8-0.42) rectangle (7, 8+0.42);
  \fill[pBfill] (7, 8-0.42) rectangle (8, 8+0.42);
\end{scope}
\draw[span, draw=black!70] (6, 8-0.42) rectangle (8, 8+0.42);
\node[spantext] at (7, 8) {MN \ldots};
\end{scope}

% ---------------- (b) the grid ----------------
\begin{scope}[yshift=-168pt]
\node[panel] at (-25pt, 47pt) {(b) Translated Grid: one block per source instruction,};
\node[panel] at (-25pt, 36pt) {fall-through chaining when possible};
\foreach \i/\a/\b/\s in {0/00/AB CD/pA, 1/01/CD EF\\GH/pB, 2/02/EF/pA, 3/03/GH IJ/pA, 4/04/IJ KL/pB, 5/05/KL/pA} {
  \node[cell, \s] (c\i) at (\i*27pt, 0) {\b};
  \node[hdr, below=1.5pt of c\i.south] {\a};
}
% MN is reached by both streams: half of each color
\node[cell] (c6) at (6*27pt, 0) {MN \ldots};
\begin{scope}
  \clip[rounded corners=1.5pt] (c6.north west) rectangle (c6.south east);
  \fill[pAfill] (c6.north west) rectangle (c6.south);
  \fill[pBfill] (c6.north) rectangle (c6.south east);
\end{scope}
\node[cell, draw=black!70] at (c6.west) {MN \ldots};
\node[hdr, below=1.5pt of c6.south] {06};
% fall-through edges as arcs over the row
\newcommand{\hop}[2]{\draw[arr] (c#1.north) to[out=60, in=120, looseness=0.9] (c#2.north);}
\hop{0}{2} \hop{1}{4} \hop{2}{3} \hop{3}{5} \hop{4}{6} \hop{5}{6}
% any address may be a target
\node[lbl] (ind) at (2.5*27pt, -38pt) {indirect branch, call, or return:\\ any offset is a landing site};
\foreach \t in {0,...,6} { \draw[darr] (ind.north) -- ([yshift=-10pt]c\t.south); }
\end{scope}
\end{tikzpicture}}
  \caption{(a)~Superset disassembly decodes an instruction at every byte offset.
  (b)~One block per offset, each falling through to the block at offset plus length. Colors mark the two fall-through streams.}
  \label{fig:superset-grid}
\end{figure}

\coolname{} uses a form of superset disassembly~\cite{Bauman2018SupersetDS} to create a ``grid of unitary basic blocks'' corresponding to every possible jump target address in the original program text. Since x86 has a variable-length instruction format and branches may potentially land in the ``middle'' of an instruction, this means that we potentially need to re-interpret the constituent bytes of such instructions multiple times (Figure~\ref{fig:superset-grid}a), generating a separate corresponding pre-compiled basic block in the target instruction set for every byte offset in the input binary. Each of these pre-compiled basic blocks implements just a single instruction of the original program and then branches to the translated basic block corresponding to the instruction that follows in the original program (Figure~\ref{fig:superset-grid}b). Since we explicitly and accurately model all the flag computations of the original instruction, and we potentially create multiple interpretations for each byte of the input text segment, our statically translated binaries are much larger than the corresponding input binaries and a purely ``grid-based'' execution on the target platform would be slow.

% Hence, we subsequently profile the resulting ``grid-based'' output binary on the target platform to observe actual branch targets and use the accumulated profiling information to add additional, optimized ``confirmed'' control-flow paths to the existing grid. This process is different from any previous compilation system that we are aware of. In our approach, we start with a ``grid'' of pre-compiled paths for every byte address in the input binary. As a consequence, compiling side-exits from profile-guided optimized program traces becomes trivial: every possible target address in the address space of the input text segment has a corresponding translated address in the output binary by construction. The end result of adding optimized confirmed path overlays to the basic grid structure is an even larger binary than before, but one that performs surprisingly well.

Hence, we subsequently profile the resulting ``grid-based'' output binary on the target platform to observe actual branch targets and add optimized ``confirmed paths'' to the existing grid. This overall process is different from any previous compilation system that we are aware of. In our approach, we start with a ``grid'' of pre-compiled paths for every byte address in the input binary. As a consequence, compiling side-exits from confirmed paths becomes trivial: every possible target address in the address space of the input text segment has a corresponding translated address in the output binary by construction. The end result of adding optimized confirmed paths to the basic grid structure is an even larger binary than before, but one that performs surprisingly well.

Maintaining the grid in addition to the confirmed paths allows \coolname{} to handle seamlessly any input binary, including obfuscated ones. Obfuscation techniques such as mixed boolean-arithmetic (MBA)~\cite{mba} and virtualization~\cite{vmprotect,themida,tigressPaper} destroy structural patterns that can provide hints about where indirect control flow may go. In the real world, binaries are often not ``well-behaved'' and feature oddities such as hand-written assembly. We use obfuscation as a controllable stand-in for these oddities, as both fall outside what standard compilers emit and what a heuristic-based binary translator can handle. As we demonstrate in this paper, our approach can trivially rewrite these oddities, including even ones that rely on overlapping instruction sequences. We expect that this robustness will also extend to other oddities in real-world binaries. Note that our system's current inability to handle C++ exceptions and multi-threaded binaries is not due to a fundamental limitation of our translation technique. Exception handling depends on the specific implementation of stack unwinding in ELF binaries, which will require significant implementation effort to replicate without yielding proportional scientific insights. Multi-threaded translation requires bridging the strong x64 memory model to the weak AArch64 one, a problem that Lasagne, Risotto, and Arancini have already addressed for static, dynamic, and hybrid translation respectively~\cite{rocha2022lasagne,risotto,arancini2026}. Their fence insertion works on the loads and stores of any translated program, so we expect it to compose with our approach.
%and we leave it out of this paper.
In summary, this paper makes the following principal contributions:
\begin{itemize}
%\item To the best of our knowledge, 
\item \coolname{} is the first fully-static binary translator that combines a ``grid'' of statically translated instructions for every byte offset in the original binary with ``confirmed paths'' of optimized control flows based on statically discovered branch targets and observation-guided path selection (OGPS). The grid by itself provides the entire functionality of the input program on the target architecture and is constructed first. Optimizations can be aggressive and are added as overlays on top of the existing grid. We place ``hot'' code in the main ELF segment and ``cold'' code in a \texttt{.text.unlikely} ELF segment.
\item The resulting system architecture is simpler than state of the art hybrid dynamic compilers, but yet consistently generates code that performs at least as well. In effect, we are presenting a novel method of trading code size for better execution performance.
\item As a side effect of our approach, our system can trivially handle even heavily obfuscated code.
\end{itemize}
% We have no financial interest in this work. We will gladly participate in the artifact evaluation process and will open-source our software at the end of our project.
%All parts of \coolname{} will be released as open source at the end of the project, and we will submit the system for artifact evaluation.

\section{System Overview}
\label{overview}
\coolname{} consumes arbitrary x86-64\footnote{We have implemented our system for x86-64 input binaries. For better readability, we abbreviate this as ``x64'' in most of the text.} ELF binaries, utilizing only ELF metadata required by the loader (i.e., no debugging information), and produces corresponding standalone AArch64 ELF output executables. It does so statically (``ahead of time'') without using just-in-time compilation, without interpreter fallback, and without requiring any kind of special runtime support library or operating system component on the target platform.

Our binary translation comprises two distinct phases. In a first phase, we decode the text section of the input program using a form of superset disassembly. Individual x64 instructions have variable lengths and can span up to 15 bytes in total. Instead of statically deciding whether a specific byte in the text segment is code or data or trying to find the beginning of each instruction in the byte stream, we use every possible interpretation of every byte in the input binary simultaneously. To this effect, we start a new decode from every byte offset in the input text segment until we have parsed exactly one instruction (of variable length) (Figure~\ref{fig:superset-grid}a). We then translate that one instruction into its own ``unitary'' basic block that contains the functionality represented by the translated instruction, any required flag computations, and finally a branch to the basic block that corresponds to the instruction that follows the original instruction in the original program (Figure~\ref{fig:superset-grid}b); if the instruction is a branch, its control flow will be translated into separate \emph{taken} and \emph{fall-through} branches.

The resulting code structure resembles a grid: a dense mesh of short basic blocks with lots of branches among them, reflecting the fact that the execution of consecutive instructions in the input program will often need to be realized by branching in the translated program. Naturally, each such grid also contains lots of "superset noise" where data embedded within the program or operands of other instructions were translated as instructions, despite never being reached during actual execution. However, the advantage of the ``grid'' representation is that for every possible branch target address within the range of the input text segment, we have created a corresponding translation a priori. That means that we don't need to perform any sophisticated analysis to ensure that we have captured all possible computed branch targets, and we don't need any kind of fallback mechanism to deal with any potential ``new'' branch targets that are discovered ``late'' during execution. The ``grid'' is a complete representation of the input program---in spite of the fact that disassembly is statically undecidable.\footnote{As Horspool showed in 1980~\cite{horspoolAndMarovac1980detranslation}, for most computer architectures, the general problem of decompiling or disassembly is equivalent to the Halting Problem~\cite{turing1967halting}.}

In a second phase, we then construct ``confirmed'' control flow paths by discovering and stitching together sequentially executed instructions to form longer basic blocks. These paths start at the program entry point, at the branch targets observed during profiling, and at the not-taken sides of branches observed during profiling; from the entry point and the not-taken sides, further control flow is traced statically. Each confirmed path we create replaces an existing node in the grid that corresponds to the first instruction in the path. If control ever ends up targeting this instruction during execution, it is now on a confirmed path.
%frequently executed traces. In contrast to previous hybrid or dynamic compilation engines, we are constructing these optimized control-flow paths on top of the ``grid'' structure that all by itself already represents an embodiment of the entire, fully translated input program. As a consequence, we have an existing defined translated code address for every byte offset in the input binary. This greatly simplifies the use of aggressive optimization for individual control-flow traces, since at any point we can bail out to a defined location in the grid without requiring de-optimization.

% Single-column version of the pipeline overview. Stage 2 runs top to
% bottom; Stage 1 (the semantics bank) feeds instantiation from the right; Stage 3
% takes the executable, runs it under perf + SPE, and feeds the profile
% back into chain extension.
\begin{figure}[h]
\centering
\begin{tikzpicture}[
  font=\footnotesize,
  >={Stealth[length=1.6mm,width=1.3mm]},
  box/.style ={draw, rounded corners=2pt, align=center,
               inner xsep=3pt, inner ysep=3pt, minimum height=7mm, text width=30mm},
  io/.style  ={box, fill=black!5},
  proc/.style={box, fill=blue!7},
  lib/.style ={box, fill=orange!12},
  ovcap/.style={font=\footnotesize\itshape, align=center},
  flow/.style={->, thick},
  feed/.style={->, thick, dashed},
  node distance=3mm,
]

% ----- Stage 2: per-binary translation, top to bottom -----
\node[io]   (x86)                        {x86-64 ELF binary};
\node[proc] (disasm) [below=of x86]      {Superset disassembly\\$\to$ grid of blocks};
\node[proc] (tiles)  [below=of disasm]   {Semantics instantiation\\per grid block};
\node[proc] (chain)  [below=of tiles]    {Inlining \& chain extension};
\node[proc] (emit)   [below=of chain]    {LLVM back-end \& section emission};
\node[io]   (out)    [below=of emit]     {AArch64 ELF executable};

\draw[flow] (x86)    -- (disasm);
\draw[flow] (disasm) -- (tiles);
\draw[flow] (tiles)  -- (chain);
\draw[flow] (chain)  -- (emit);
\draw[flow] (emit)   -- (out);

\begin{scope}[on background layer]
  \node[draw, dashed, rounded corners=3pt, inner sep=5pt,
        fit=(x86)(disasm)(tiles)(chain)(emit)(out)] (stage2) {};
\end{scope}
\node[ovcap, anchor=south west] at ([yshift=1pt]stage2.north west)
     {\textbf{Stage~1}: Whole executable binary translation};

% ----- Stage 3: profile feedback, beside emission and the executable -----
\node[proc] (perf) [right=9mm of out, text width=26mm]
     {Run on Arm hardware under \texttt{perf} + Arm SPE};
\node[proc] (agg)  [above=of perf, text width=26mm]
     {Aggregate SPE samples\\$\to$ profile of\\observed targets};

\draw[feed] (out.east)  -- (perf.west);
\draw[flow] (perf)      -- (agg);
\draw[feed] (agg.west)  -| ([xshift=4mm]chain.east) -- (chain.east);

\begin{scope}[on background layer]
  \node[draw, dashed, rounded corners=3pt, inner sep=5pt, fit=(agg)(perf)] (stage3) {};
\end{scope}
\node[ovcap, anchor=north west] at ([yshift=-1pt]stage3.south west)
     {\textbf{Stage~2}: Observation-guided\\selection of confirmed paths};

\end{tikzpicture}
\caption{\coolname{} pipeline.}
\label{fig:overview}
\end{figure}
To perform observation-guided path selection, we execute the ``grid'' representation on Arm hardware under \texttt{perf} with Arm SPE. The sampled branch destinations are aggregated into a profile. This selection phase can be re-iterated as necessary to add additional confirmed control-flow paths.
%(Section~\ref{sec:ogps}).

% \msof{In principle, at the cost of an enormous further code blowup, instead of using profiling, one could also exhaustively stitch together longer control-flow paths in a deterministic manner. For example, for any straight instruction sequence in the original program from any starting address in the original program one could construct a single coherent longer basic block. This somewhat invalidates Q2 and Q3.}

Our specific implementation in \coolname{} benefits from the fact that our target architecture has many more registers than our source architecture. However, our general method could also be applied if this were not the case, e.g. using emulated registers in memory, but this would undoubtedly have a detrimental effect on performance.

In our implementation, each x64 register is held in a dedicated AArch64 register throughout the entire execution of a translated binary. For every byte offset in the input that decodes to a valid x64 instruction, \coolname{} produces an LLVM IR function under a custom calling convention we implement on LLVM~22, pinning each x64 register to its assigned AArch64 register. % (Section~\ref{sec:cc}).
The function body is instantiated from a \emph{semantic template}, a set of small C functions compiled to LLVM IR that together encode the instruction's semantics. This granularity is imposed by the nature of static binary translation. Neither indirect-branch targets nor the code-versus-data boundary can be determined from disassembly alone. Later stages additionally combine fall-through sequences of these per-x64-instruction functions into larger ones, forming \emph{confirmed paths} that LLVM's standard optimization pipeline can run across multiple x64 instructions at once (Section~\ref{sec:confirmed-path}).  Optionally, an OGPS pass uses runtime branch samples to extend confirmed path coverage and refine the binary's section layout (Section~\ref{sec:ogps}).

% \msof{I think we should use the words grid and express path throughout. It may even make sense to have a picture that shows overlapping instructions being translated into a grid. I think that would enable us to completely get rid of the entire ``tile'' discussion, making this even more separate from Elevator. We simply don't say how we build the grid, we just do it.}

% \msof{Since they will only read 2 pages to begin with, we shouldn't have a background section but move all that to a related work section}

\paragraph{Translation Semantics}
\coolname{} stores the semantics of x86 instructions in a bank of LLVM bitcode built once, offline. Like Elevator~\cite{elevator}, a recent binary translator that also uses superset disassembly, it specifies the semantics as C functions. Elevator compiles these functions to AArch64 machine code and stitches the machine code together per instruction. \coolname{} instead models them as LLVM IR. %Clang turns each function into IR, and
To translate an element of the grid, the relevant functions are copied, specialized, and concatenated into one linear IR sequence. Keeping the semantics as IR is what lets LLVM optimize across instruction boundaries when building confirmed paths, which matters most for eliminating dead flag computations.

\paragraph{Calling Convention}
\label{sec:cc}
Translated grid elements must agree on where the x64 state lives during execution. Because AArch64 has more registers than x64, \coolname{} models the entire x64 register file within the AArch64 register file. Each element is compiled as a complete function under a custom calling convention taking 33 parameters, and ends in a tail call passing these 33 values to the next element. This is very similar to the stencils used by copy-and-patch~\cite{copyAndPatch}. The x64 register state is therefore pinned in AArch64 registers at every element boundary, so that an indirect branch can target any grid element at runtime and find the x64 state where it expects it.
\paragraph{Register Mapping}
\label{sec:register-map}
Each grid element's implementation function takes 33 parameters because that is how many AArch64 registers represent the x64 state. The 16 x64 general-purpose registers and \texttt{RFLAGS} occupy 17 \texttt{X} registers, and the 16 \texttt{XMM} registers occupy 16 \texttt{V} registers. The remaining AArch64 registers are left unassigned and stay available to LLVM when lowering individual grid elements. This helps avoid spilling inside an element when evaluating complex x64 semantics. AArch64 supports fewer complex operations than x64 and often needs additional registers to express them: an x64 memory operand such as \texttt{[RAX+RDX*4+12345678h]} cannot be encoded in any AArch64 load/store instruction and must be computed with explicit arithmetic.

\paragraph{Two Stack Pointers}
\label{sec:stack-pointers}
\coolname{} maps the x64 stack pointer like every other register as one of the 33 arguments, pinned to AArch64's \reg{X15}. \coolname{} does not emulate the x64 stack in separate memory. The translated program runs on the ordinary process stack, so the x64 stack state is shared with AArch64's \reg{SP}.

Many x64 instructions modify the stack pointer in ways that cannot be lowered directly onto AArch64's \reg{SP}. AArch64 checks stack pointer alignment in hardware, faulting on any load or store that uses \reg{SP} as its base register unless \reg{SP} is 16-byte aligned. x64 has no such check. Its ABI expects 16-byte alignment at call sites purely as a software convention, so stack accesses that are legal on x64 can fault on AArch64. \coolname{} therefore performs all x64 stack operations in \reg{X15}, where no alignment rule applies. The actual \reg{SP} register is brought up to date from \reg{X15} before it is next used.

\begin{figure}[h]
  \centering
  \adjustbox{max width=\columnwidth}{% The stack of a translated function that spills, as a picture instead of
% the prologue/epilogue listing. Addresses grow upward. X15 holds the x64
% stack pointer (RSP) and is never moved by the frame; SP is derived from
% it on entry and restored from it on exit, so the function enters and
% leaves with SP == X15. The 128-byte red zone below RSP belongs to the
% x64 code and must not be touched; the spill slots sit below it, at a
% 16-byte aligned SP so that SP-based loads and stores do not fault.
% Fonts are \footnotesize (8pt) to meet the ASPLOS figure-text minimum.
\definecolor{inkPrimary}{HTML}{2b2b28}
\definecolor{inkMuted}{HTML}{898781}
\definecolor{stkFill}{HTML}{e1e0d9}\definecolor{stkEdge}{HTML}{898781}
\definecolor{redFill}{HTML}{f2d3cf}\definecolor{redEdge}{HTML}{a8524a}
\definecolor{spFill}{HTML}{b9cbdd}\definecolor{spEdge}{HTML}{41638a}
\begin{tikzpicture}[
  x=1pt, y=1pt,
  seg/.style={line width=0.4pt, font=\ttfamily\footnotesize, text=inkPrimary, align=center, inner sep=1pt},
  stk/.style={seg, draw=stkEdge, fill=stkFill},
  rzone/.style={seg, draw=redEdge, fill=redFill},
  pad/.style={seg, draw=stkEdge, fill=white},
  spill/.style={seg, draw=spEdge, fill=spFill},
  ptr/.style={-{Stealth[length=4pt]}, line width=0.8pt},
  lbl/.style={font=\ttfamily\footnotesize, inner sep=1pt, align=left},
  note/.style={font=\ttfamily\footnotesize, text=inkMuted, inner sep=1pt, align=left},
]
% ---- memory column: high addresses at the top ----
\def\W{96}      % width of the memory column (fits "+ const stack adjust" at 8pt)
\node[stk,   minimum width=\W pt, minimum height=16pt, anchor=south west] (stk)  at (0, 78) {x64 stack};
\node[pad,   minimum width=\W pt, minimum height=12pt, anchor=north west] (pad)  at (0, 78) {align pad};
\node[rzone, minimum width=\W pt, minimum height=22pt, anchor=north west] (rz)   at (pad.south west) {red zone, 128 B\\+ const stack adjust};
\node[spill, minimum width=\W pt, minimum height=16pt, anchor=north west] (sp)   at (rz.south west) {spill slots};
\node[note, anchor=north] at ([yshift=-2pt]sp.south) {lower addresses};
\node[note, anchor=south] at ([yshift=2pt]stk.north) {higher addresses};

% ---- pointers on the left ----
\draw[ptr, redEdge] (-30, 78) -- (stk.south west);
\node[lbl, anchor=east, text=redEdge] at (-32, 78) {X15};
\draw[ptr, spEdge] (-30, {78-12-22-16}) -- (sp.south west);
\node[lbl, anchor=east, text=spEdge] at (-32, {78-12-22-16}) {SP};

% ---- what the prologue and epilogue do, on the right ----
\node[note, anchor=west] at ([xshift=4pt]rz.east) {entry: X15 = SP,\\then lower SP};
\node[note, anchor=west] at ([xshift=4pt]sp.east) {exit: SP = X15};
\node[note, anchor=west] at ([xshift=4pt]stk.east) {x64 stack ops\\use X15};
\end{tikzpicture}}
  \caption{Stack of a translated grid element that spills.}
  \label{fig:frame}
\end{figure}
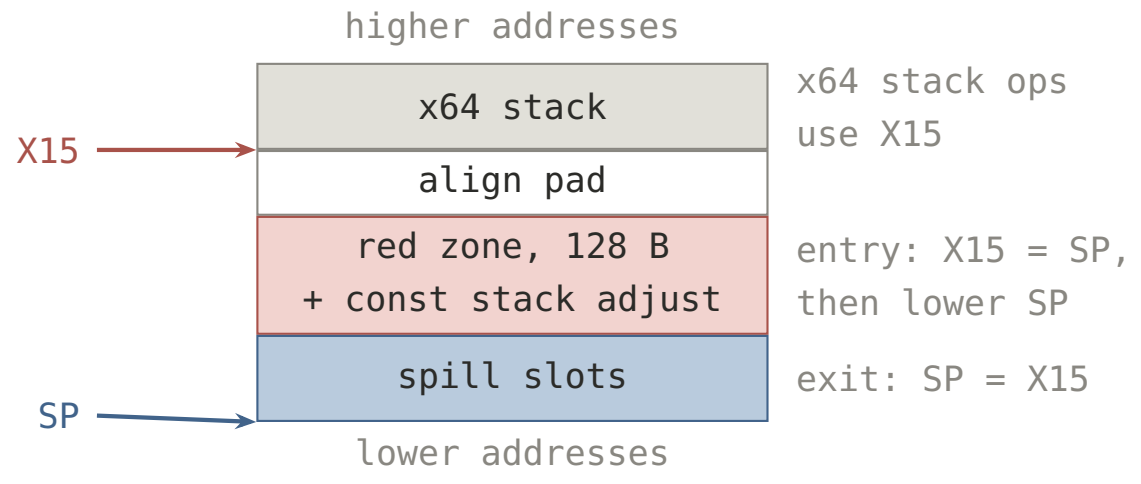

A grid element can also need the stack even when its x64 instructions never touch \reg{RSP}. When live values outnumber the unpinned registers, LLVM spills the excess. This spill space cannot sit directly below \reg{SP}, because the System~V ABI reserves the 128 bytes below \reg{RSP} as a red zone that the translated program may already be using. Since \reg{SP} tracks \reg{RSP}, spilling there would clobber those values. Figure~\ref{fig:frame} shows how we lower the spill slots to make sure there is no overlap.

After the LLVM generated epilogue, a single instruction moves \reg{X15} into \reg{SP}. If the grid element modified \reg{RSP}, this installs the new value; otherwise it restores the pointer to its state on entry.

\paragraph{ABI Translation} 
While control remains inside the translated binary, the original x64 ABI is upheld, but in the AArch64 registers assigned by our mapping. When control enters or leaves the binary, the emulated x64 ABI must be translated into the AArch64 Procedure Call Standard that external functions expect. We adopt the techniques introduced by Elevator~\cite{elevator}: indirect branch and call instructions are instrumented to detect exits, and a signal handler detects re-entries.

\paragraph{Flag Liveness}                    
\label{sec:eflags}     
%The x64 \texttt{RFLAGS} register holds six status bits: carry, overflow, zero, sign, parity, and auxiliary. AArch64 keeps four condition flags in \texttt{PSTATE}: negative (N), zero (Z), carry (C), and overflow (V), corresponding to the x64 sign, zero, carry, and overflow bits. Parity and auxiliary have no counterpart and must be emulated.

Computing all six bits after every flag-writing instruction is expensive and unnecessary in many cases. If a flag is not read before being written by a post-dominating instruction, its semantics can be entirely omitted from the grid. 

Additionally, when a conditional branch reads only the carry, overflow, zero, or sign flags and follows the operation that computes them in a confirmed path, \coolname{} uses the native AArch64 \texttt{PSTATE} flags instead. The comparison is emitted as an LLVM \texttt{icmp} that the branch or conditional consumes directly, leaving the AArch64 backend to select the corresponding condition code rather than materializing flag values.
%A flag function's result matters only if a later instruction reads that bit before another instruction overwrites it. \coolname{} therefore computes flag liveness for every decoded instruction by looking forward at post-dominating instructions in the grid.
%in a single backward walk over the fall-through order, maintaining the set of bits live at the current point. At each instruction it records which of the bits that instruction writes are still live. Only the flag functions for those bits are cloned into the translated function.

% Single-column, dense version of pipeline-listings.tex: the running example
% at four stages, stacked. Panel labels keep the former listing labels
% (lst:tile, lst:pertile, lst:pertile-inlined, lst:chain). Long IR lines are
% wrapped by hand so nothing exceeds the column at 8pt mono (60 chars).
% Content still shows sub ax,0x84 @0x85199; swap in the new running example
% (XXH64 loop, 777.zstd @0x186ec) once the IR is regenerated.
\begin{figure}[t]
\captionsetup[subfigure]{font=scriptsize, skip=1pt, aboveskip=1pt, belowskip=3pt}
\lstset{basicstyle=\fontsize{8}{9}\selectfont\ttfamily, columns=fixed,
  basewidth=0.5em, xleftmargin=0pt, framexleftmargin=0pt, aboveskip=0pt,
  belowskip=0pt, lineskip=-0.5pt}
% \begin{subfigure}{\columnwidth}
% \begin{lstlisting}[language=C]
% uint64_t SUB16_R1_R1_R2(uint64_t R1, uint64_t R2) {
%   return ((R1 - R2) & MASK16) | (R1 & ~MASK16); }
% uint64_t SUB16_ZF_R1_Torg_R2_FLAGS(uint64_t R1,
%     uint64_t Torg, uint64_t R2, uint64_t FLAGS) {
%   FLAGS &= ~FLAG_ZF;
%   if ((R1 & MASK16) == 0) FLAGS |= FLAG_ZF;
%   return FLAGS; }
% \end{lstlisting}
% \caption{Semantics bank: two of the seven C functions for 16-bit
% register subtraction, the destination value and the zero flag.}
% \label{lst:tile}
% \end{subfigure}

\begin{subfigure}{\columnwidth}
\begin{lstlisting}[language=LLVM,
  emph={[4]@x86_85199,@x86_8519D,@SUB16_R1_R1_IMM,
    @SUB16_CF_R1_Torg_IMM_FLAGS,@SUB16_ZF_R1_Torg_IMM_FLAGS},
  emphstyle={[4]\color{llvmfn}},
  emph={[5]\%rax,\%rcx,\%rflags,\%ax,\%f1,\%f2,\%fl},
  emphstyle={[5]\color{llvmvar}}]
define cc 128 void @x86_85199(i64 %rax, ...,
                              i64 %rflags, ...) {
  %ax = call i64 @SUB16_R1_R1_IMM(i64 %rax, i64 132)
  %f1 = call i64 @SUB16_CF_R1_Torg_IMM_FLAGS(
            i64 %ax, i64 %rax, i64 132, i64 %rflags)
  %f2 = call i64 @SUB16_ZF_R1_Torg_IMM_FLAGS(
            i64 %ax, i64 %rax, i64 132, i64 %f1)
  ; ... four more flag functions, yielding %fl ...
  musttail call cc 128 void @x86_8519D(
            i64 %ax, ..., i64 %fl, ...)
  ret void }
\end{lstlisting}
\caption{The translated function for \texttt{SUB AX,0x84} at
\texttt{0x85199}, as constructed.}
\label{lst:pertile}
\end{subfigure}

\begin{subfigure}{\columnwidth}
\begin{lstlisting}[language=LLVM,
  emph={[4]@x86_85199,@x86_8519D},
  emphstyle={[4]\color{llvmfn}},
  emph={[5]\%rax,\%rcx,\%rflags,\%ax,\%diff,\%low,\%keep,
    \%z,\%zf,\%fl0,\%fl},
  emphstyle={[5]\color{llvmvar}}]
define cc 128 void @x86_85199(i64 %rax, ...,
                              i64 %rflags, ...) {
  %diff = sub i64 %rax, 132
  %low  = and i64 %diff, 65535
  %keep = and i64 %rax, -65536
  %ax   = or i64 %keep, %low
  %z    = icmp eq i64 %low, 0
  %zf   = select i1 %z, i64 64, i64 0
  %fl0  = and i64 %rflags, -65
  %fl   = or i64 %fl0, %zf
  ; ... five more flag bits ...
  musttail call cc 128 void @x86_8519D(
            i64 %ax, ..., i64 %fl, ...)
  ret void }
\end{lstlisting}
\caption{The same function after mandatory inlining, before flag pruning.}
\label{lst:pertile-inlined}
\end{subfigure}

\begin{subfigure}{\columnwidth}
\begin{lstlisting}[language=LLVM,
  emph={[4]@x86_85199,@x86_851C7,@x86_851A3},
  emphstyle={[4]\color{llvmfn}},
  emph={[5]\%rax,\%rcx,\%rflags,\%ax,\%diff,\%low,\%keep,\%cond},
  emphstyle={[5]\color{llvmvar}}]
define cc 128 void @x86_85199(i64 %rax, ...) {
  %diff = sub i64 %rax, 132        ; sub
  %low  = and i64 %diff, 65535
  %keep = and i64 %rax, -65536
  %ax   = or i64 %keep, %low
  %cond = icmp ugt i64 %low, 1     ; cmp + ja, fused
  br i1 %cond, label %taken, label %fall
taken: musttail call cc 128 void @x86_851C7(i64 %ax, ...)
       ret void
fall:  musttail call cc 128 void @x86_851A3(i64 %ax, ...)
       ret void }
\end{lstlisting}
\caption{The chain of Figure~\ref{fig:candidate-confirmed-paths}b as one LLVM
function.}
\label{lst:chain}
\end{subfigure}
\caption{One instruction at three stages.}
\label{fig:pipeline-listings}
\end{figure}

\paragraph{Per-Instruction Construction}
\label{sec:pertile}
A single instruction's translation semantics begin life as a collection of LLVM IR functions in a bank. When translating a new grid element, a single new LLVM IR function is created named for the instruction's location in the original x64 binary. Based on the x64 instruction, in this case \texttt{SUB AX,0x84}, the relevant semantics are selected from the bank and calls to them are added to the newly created IR function. This can be seen in Figure~\ref{lst:pertile}. During this initial construction, flag computations that are known to be dead are not added.

Every call in the function is marked for mandatory inlining. LLVM inlines the functions into the body, and Listing~\ref{lst:pertile-inlined} shows the result: straight-line IR representing the semantics of the x64 instruction, but now as a single LLVM IR function.

The closing tail call of every element is handled in two ways. 
On the candidate path (see Section~\ref{sec:candidate-path}), the tail call compiles to a direct branch. On the confirmed path (see Section~\ref{sec:confirmed-path}), the translator continuously inlines successor calls into the current one until an x64 branch instruction is reached. Figure~\ref{fig:candidate-confirmed-paths} illustrates how the confirmed path can enable complete omission of flag operations.

\subsection{Laying Out ``The Grid'' --- Candidate Path Fall-Through}
\label{sec:candidate-path}
The Grid is composed of unitary basic blocks that each represent a single instruction in the input program. Conceptually, we have an explicit transfer to the next basic block (representing the next instruction in the original program) at the end of each such block. In an attempt to reduce branching overhead and increase branch predictability, we try to lay out the unitary basic blocks in the output binary in such a manner that unnecessary branching can be avoided by placing the blocks of successor instructions directly after one another and using fall-through instead of branches to transition between them. 

Placing the translated functions in the output binary raises a question a conventional compiler never faces. \coolname{} does not know where the basic blocks of the original program begin. Superset disassembly~\cite{Bauman2018SupersetDS} produces a translated function for every valid byte offset instruction, so real instructions and phantom decodes sit interleaved in the address space, and separating them would require assumptions about the original binary layout. \coolname{} emits translated code in the order of the original instruction addresses. It takes the lowest address not yet placed and walks the decode's fall-through successors, laying each function immediately after its predecessor until the walk reaches code that is already placed, then starts a new run at the next unplaced address. 

\begin{figure}[t]
  \centering
  \definecolor{inkPrimary}{HTML}{2b2b28}
\definecolor{inkMuted}{HTML}{898781}
\definecolor{pAfill}{HTML}{b9cbdd}\definecolor{pAedge}{HTML}{41638a}
\definecolor{pBfill}{HTML}{c3d2b8}\definecolor{pBedge}{HTML}{4a6845}
\definecolor{flFill}{HTML}{f2d3cf}\definecolor{flEdge}{HTML}{a8524a}
\begin{tikzpicture}[
  hdr/.style={font=\ttfamily\footnotesize, text=inkMuted, inner sep=0},
  cell/.style={rounded corners=1.5pt, line width=0.4pt,
    font=\ttfamily\footnotesize, text=inkPrimary,
    minimum width=22pt, minimum height=13pt, inner sep=1pt, anchor=west},
  pA/.style={cell, draw=pAedge, fill=pAfill},
  pB/.style={cell, draw=pBedge, fill=pBfill},
  ghost/.style={cell, draw=pAedge, densely dashed, fill=pAfill!45!white, text=black!60},
  tab/.style={draw=flEdge, fill=flFill, line width=0.4pt},
  tabghost/.style={tab, densely dashed, fill=flFill!45!white},
  chain/.style={draw=pAedge, fill=pAfill, rounded corners=1.5pt, line width=0.8pt,
    font=\ttfamily\footnotesize, text=inkPrimary, minimum width=120pt, minimum height=13pt,
    inner sep=2pt, anchor=west},
  arr/.style={-{Stealth[length=4pt]}, black!70, line width=0.5pt},
  path/.style={-{Stealth[length=4pt]}, pAedge, line width=0.9pt},
  pathB/.style={-{Stealth[length=4pt]}, pBedge, line width=0.9pt},
  absorb/.style={densely dotted, black!50, line width=0.5pt},
  ghostarr/.style={-{Stealth[length=4pt]}, pAedge!50!white, line width=0.9pt},
  merge/.style={-, rounded corners=2pt},
  entry/.style={-{Stealth[length=4pt]}, black!70, line width=0.9pt},
  lbl/.style={font=\ttfamily\footnotesize, text=black!55, align=center, inner sep=1pt},
  panel/.style={font=\ttfamily\footnotesize\bfseries, text=inkPrimary, anchor=west, inner sep=0},
]
% Cell #1 of style #2 at (#3, #4) with text #5.
\newcommand{\cellat}[5]{\node[#2] (#1) at (#3, #4) {#5};}
% Flag tab on the right edge of cell #1, style #2; names its east edge #1tab.
% (No text inside: the red tab itself is the marker, as the caption says.)
\newcommand{\flags}[2]{%
  \draw[#2] (#1.north east) rectangle ([xshift=7pt]#1.south east);
  \coordinate (#1tab) at ([xshift=7pt]#1.east);}
% 06: reached by both streams, half of each color.
\newcommand{\gcell}[3]{%
  \node[cell] (#1) at (#2, #3) {MN \ldots};
  \begin{scope}
    \clip[rounded corners=1.5pt] (#1.north west) rectangle (#1.south east);
    \fill[pAfill] (#1.north west) rectangle (#1.south);
    \fill[pBfill] (#1.north) rectangle (#1.south east);
  \end{scope}
  \node[cell, draw=black!70] at (#1.west) {MN \ldots};}

% ================= (a) superset =================
\node[panel] at (-24pt, 36pt) {(a) Superset: each flag writer computes its bits (red)};
\cellat{a0}{pA}{-24pt}{0}{AB CD}    \flags{a0}{tab}
\node[pB] (a1) at ([xshift=3pt]a0tab) {CD EF GH};
\node[pA] (a2) at ([xshift=3pt]a1.east) {EF};      \flags{a2}{tab}
\node[pA] (a3) at ([xshift=3pt]a2tab) {GH IJ};   \flags{a3}{tab}
\node[pB] (a4) at ([xshift=3pt]a3tab) {IJ KL};
\node[pA] (a5) at ([xshift=3pt]a4.east) {KL};
\node[cell] (a6) at ([xshift=3pt]a5.east) {MN \ldots};
\begin{scope}
  \clip[rounded corners=1.5pt] (a6.north west) rectangle (a6.south east);
  \fill[pAfill] (a6.north west) rectangle (a6.south);
  \fill[pBfill] (a6.north) rectangle (a6.south east);
\end{scope}
\node[cell, draw=black!70] at (a6.west) {MN \ldots};
\foreach \i/\a in {0/00,1/01,2/02,3/03,4/04,5/05,6/06} {
  \node[hdr, anchor=north] at ([yshift=-2pt]a\i.south) {\a}; }
\newcommand{\hopa}[3]{\draw[#3] (a#1.north) to[out=60, in=120, looseness=0.9] (a#2.north);}
\hopa{1}{4}{pathB} \hopa{4}{6}{pathB}
\hopa{0}{2}{path} \hopa{2}{3}{path} \hopa{3}{5}{path} \hopa{5}{6}{path}
\draw[path] ([xshift=8pt]a5.south) -- ++(0, -14pt) node[lbl, anchor=north] {taken};

% ================= (b) candidate path =================
\begin{scope}[yshift=-72pt]
\node[panel] at (-24pt, 22pt) {(b) Candidate path: one function per block, dead bits dropped};
% stream from 00, contiguous in memory: 00 02 03 05 06, touching
\cellat{b0}{pA}{0pt}{0}{AB CD}
\node[pA] (b2) at (b0.east) {EF};
\node[pA] (b3) at (b2.east) {GH IJ};  \flags{b3}{tab}
\node[pA] (b5) at (b3tab) {KL};
\foreach \i/\a in {0/00,2/02,3/03,5/05} {
  \node[hdr, anchor=south] at ([yshift=2pt]b\i.north) {\a}; }
\draw[path] ([xshift=-14pt]b0.west) -- (b0.west);
\node[lbl, anchor=south east] at ([yshift=1pt]b0.west) {from 00};
\draw[path] (b5.south) -- ++(0, -8pt) node[lbl, anchor=north west, xshift=-3pt] {taken};
% stream from 01, contiguous: 01 04, touching; then a jump into 06
\cellat{b1}{pB}{12pt}{-32pt}{CD EF GH}
\node[pB] (b4) at (b1.east) {IJ KL};
\foreach \i/\a in {1/01,4/04} {
  \node[hdr, anchor=south] at ([yshift=2pt]b\i.north) {\a}; }
\node[lbl, anchor=east] at ([xshift=-2pt]b1.west) {from 01};
% 06: a join, apart from both streams; both paths merge into its start
\gcell{b6}{150pt}{-16pt}
\node[hdr, anchor=south] at ([yshift=2pt]b6.north) {06};
\draw[path]  (b5.east) to[out=0, in=180, looseness=1.1] ([yshift=3pt]b6.west);
\draw[pathB] (b4.east) to[out=0, in=180, looseness=1.1] ([yshift=-3pt]b6.west);
\end{scope}

% ================= (c) confirmed path =================
\begin{scope}[yshift=-158pt]
\node[panel] at (-24pt, 22pt) {(c) Confirmed path: one chain, no boundary inside};
\node[chain] (ch) at (0pt, 0) {x86\_00:\quad AB CD\,\(\cdot\)\,EF\,\(\cdot\)\,GH IJ\,\(\cdot\)\,KL};
\draw[path] ([xshift=-14pt]ch.west) -- (ch.west);
\node[lbl, anchor=south east] at ([yshift=1pt]ch.west) {from 00};
\draw[path] ([xshift=-8pt]ch.south east) -- ++(0, -9pt) node[lbl, anchor=north west, xshift=-3pt] {taken};
% absorbed standalones kept, dashed, under the chain
\cellat{c2}{ghost}{24pt}{-27pt}{EF}
\node[ghost] (c3) at (c2.east) {GH IJ};  \flags{c3}{tabghost}
\node[ghost] (c5) at (c3tab) {KL};
\foreach \i in {2,3,5} { \draw[absorb] (c\i.north) -- (c\i.north |- ch.south); }
\node[lbl, anchor=east] at ([xshift=-3pt]c2.west) {standalones\\kept};
% the stream from 01 is unchanged
\cellat{c1}{pB}{12pt}{-54pt}{CD EF GH}
\node[pB] (c4) at (c1.east) {IJ KL};
\node[lbl, anchor=east] at ([xshift=-2pt]c1.west) {from 01};
% 06: a join, apart; the chain's not-taken exit and the other stream merge into it
\gcell{c6}{170pt}{-20pt}
\node[hdr, anchor=south] at ([yshift=2pt]c6.north) {06};
\draw[path]  (ch.east) to[out=0, in=180, looseness=1.1] ([yshift=3pt]c6.west);
\draw[pathB] (c4.east) to[out=0, in=180, looseness=1.1] ([yshift=-3pt]c6.west);
\end{scope}
\end{tikzpicture}
\caption{(a) Individually translated instructions, (b) candidate path eliminating dead flag writes laid out for fall-through, and (c) fully optimized confirmed path for the same input.}
  \label{fig:candidate-confirmed-paths}
\end{figure}
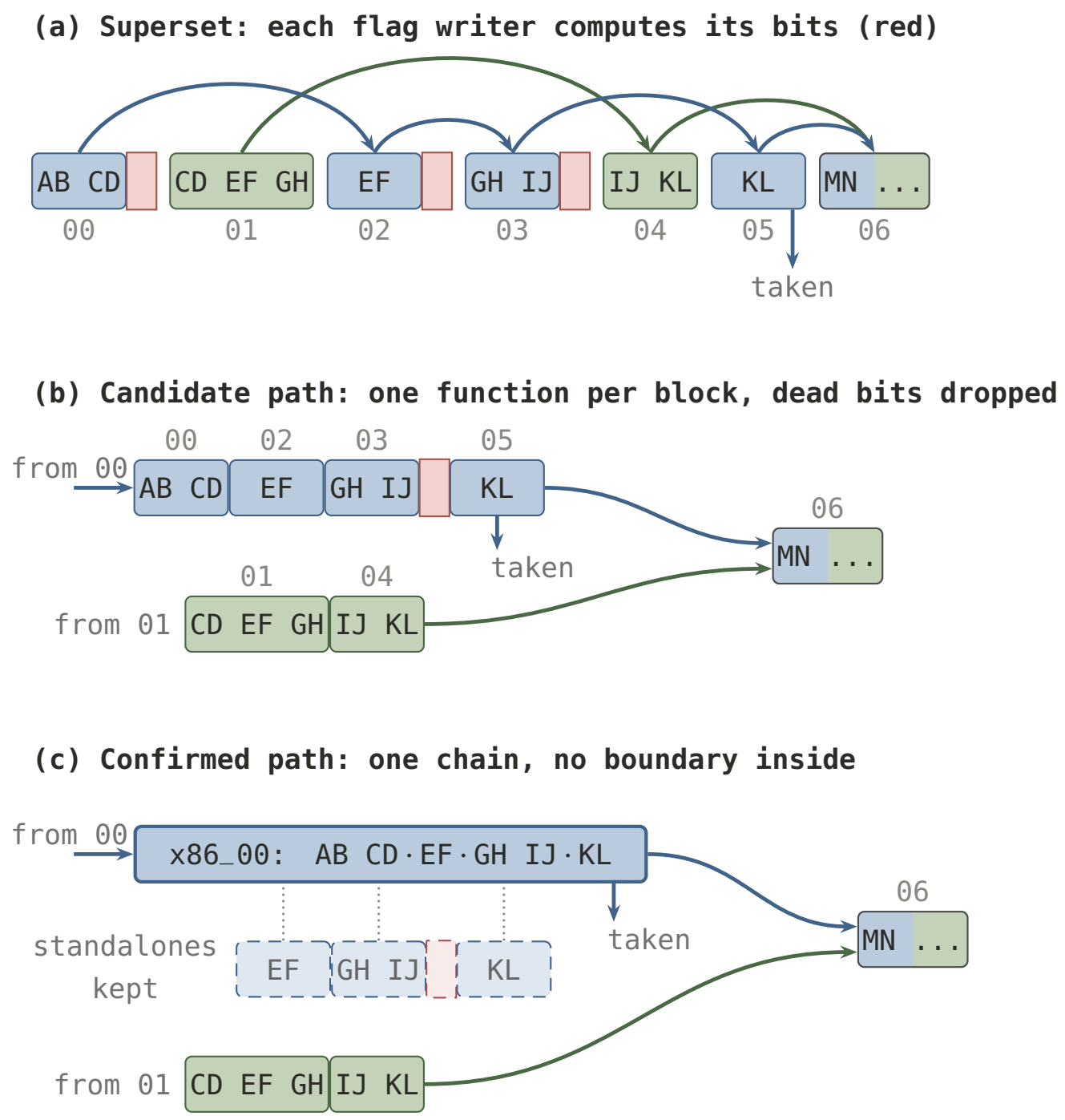

We call the resulting layout the candidate path (Figure~\ref{fig:candidate-confirmed-paths}b). Candidate paths are paths that are correctly translated on their own and do not involve any optimization with the instructions before and after, other than dropping dead flag writes. Execution proceeds one translation after another, in the original program order. %, crossing a function boundary at every instruction.
The candidate path is the correctness substrate of the translation. Every x64 address has a translated function at a known location, so any control transfer, including an indirect branch whose target no static analysis could predict, has a correct location to land. Adjacent translations on the candidate path stay independent of one another. Each is compiled individually % LLVM optimizes only within one translation
%\msof{that is not correct - what about dead flag write elimination?}
and all 33 register values are pinned to their assigned registers at every boundary. The candidate path is therefore complete but pays a boundary cost when transitioning between translated instructions.

\subsection{Confirmed Path - Chain Construction}
\label{sec:confirmed-path}
On the candidate path, optimizations don't cross the boundaries of translated instructions. But of course, larger compilation units would enable further optimization opportunities. To this effect, \coolname{} contains a mechanism to automatically aggregate chains of basic blocks into larger compilation units. Any branch target can be the head of such a chain. Beginning at the head, the translator inlines the translation's closing tail call with the body of its successor, until the next address is a branch instruction or the instruction itself is a branch target. The ``absorbed'' instructions form one larger translation inside a single LLVM function, which serves as a basic block and is optimized with dead-code elimination, constant propagation, etc. Confirmed paths are formed from three sources: control flow that is statically traceable from the program entry point, control flow observed during profiling, and control flow that is statically traceable from the not-taken sides of branches observed during profiling. Figure~\ref{fig:candidate-confirmed-paths}c shows a chain at \texttt{x86\_85199}, which absorbs the two subsequent instructions, and Listing~\ref{lst:chain} shows the resulting function.

The interesting question is where to start building such longer chained compilation units when our program representation is littered with phantom basic blocks due to our use of superset disassembly. In theory, one could build longer chains for every unitary basic block in the translated program, leading to an enormous further code expansion. We instead opted to use all branch targets that are statically reachable from the binary's entry points and additionally use a small amount of profiling to pinpoint dynamic branch targets seen during actual execution. This approach yields surprisingly good performance results, as shown in our evaluation.

\section{Observation-Guided Path Selection}
\label{sec:ogps}

Observation-guided path selection (OGPS) runs the translated binary, records which branch targets are reached, and selects those addresses as entries of confirmed paths. In order to find dynamic branch targets that would be worthwhile to aggregate into confirmed paths, 
\coolname{} runs the translated binary on the target machine with Arm's Statistical Profiling Extension~(SPE) sampling branches through Linux \texttt{perf}. Each sample records the source and destination of one branch. \coolname{} keeps the samples that land in a translated function, maps each destination back to its x64 address through the function's symbol, and translates the binary a second time with every observed address as an entry point to a confirmed path. The profile stores the sampled edges with their counts, but \coolname{} uses only the set of destinations. OGPS therefore differs from profile-guided optimization, which depends on how often each target is observed. A single sample adds an address, and additional runs on further inputs add their addresses to the same set. If a later run lands on an address never observed during profiling, execution will take the candidate path there, as it did before profiling.

\begin{algorithm}
\caption{Observation-guided classification and placement}
\label{alg:ogps-layout}
\footnotesize
\begin{algorithmic}[1]
\Require translated functions $F$, one per decoded offset; function entries $\mathcal{E}$; static branch targets $\mathcal{S}$; absorbed chain members $\mathcal{M}$; profile $P$, possibly empty
\Ensure a class and a section for every $f \in F$
\Procedure{Classify}{$F, \mathcal{E}, \mathcal{S}, \mathcal{M}, P$}
  \State $\mathcal{O} \gets \textsc{ObservedTargets}(P)$ \Comment{runtime landings}
  \State $\textit{Entries} \gets \mathcal{E} \cup \mathcal{S} \cup \mathcal{O}$
  \State $\textit{Alive} \gets \textsc{Reach}(\mathcal{E} \cup \mathcal{O})$ \Comment{direct control flow}
    \ForAll{$f \in F$}
    \State $a \gets \textsc{Offset}(f)$ \Comment{the decoded offset $f$ translates}
    \If{$a \in \mathcal{M}$ \textbf{and} $a \notin \textit{Entries}$}
      \State $\textit{class}(f) \gets \textsc{CandidatePath}$ \Comment{copy of a chain member}
    \ElsIf{$a \notin \textit{Alive}$}
      \State $\textit{class}(f) \gets \textsc{Unreached}$
    \ElsIf{$a \in \textit{Entries}$ \textbf{or} $P = \emptyset$}
      \State $\textit{class}(f) \gets \textsc{ConfirmedPath}$
    \Else
      \State $\textit{class}(f) \gets \textsc{CandidatePath}$ \Comment{reachable}
    \EndIf
  \EndFor
  \State place \textsc{ConfirmedPath} in \texttt{.text}, in source order
  \State place \textsc{CandidatePath}, then \textsc{Unreached}, in \texttt{.text.unlikely}
\EndProcedure
\end{algorithmic}
\end{algorithm}

\paragraph{Hot/Cold Code Placement.} The observed entry points also assist code placement. The concept of considering ``working sets'' of instructions and their effect on the instruction cache has been around for a long time~\cite{10.1145/361179.361195,profile-guided-code-positioning}. In the spirit of previous work on profile-guided code layout optimizations~\cite{10.1145/384285.379260}, we automatically separate the generated code into two distinct ELF sections: a main section that contains all of the confirmed code paths, and a much larger cold-code section that is marked as \texttt{.text.unlikely}. Potentially, this enables the operating system to never actually load the vast majority of our generated code into memory.

A reachability walk from the function entries and the observed entry points follows direct control flow and marks every translation it reaches. \coolname{} places the reached translations at entry points in the main ELF section, in the order of their addresses in the input binary, and the rest in the \texttt{.text.unlikely} section, together with the standalone copies of absorbed instructions (Algorithm~\ref{alg:ogps-layout}). Because the walk starts at entry points and follows the decode stream from each, the overlapping decodes of the same bytes, which the superset also translated, stay unreached and move to the \texttt{.text.unlikely} section as well. The separation keeps the working code set close together, so it fills fewer pages and cache lines than it would if mixed with translations that never run. The walk does not claim to know which decodes are real. It records what the entry points reach, and missed translation stay in the binary and can still be reached. Section~\ref{sec:runtime} measures the resulting split.

\begin{figure*}[t]
    \centering
    \includegraphics[width=\textwidth]{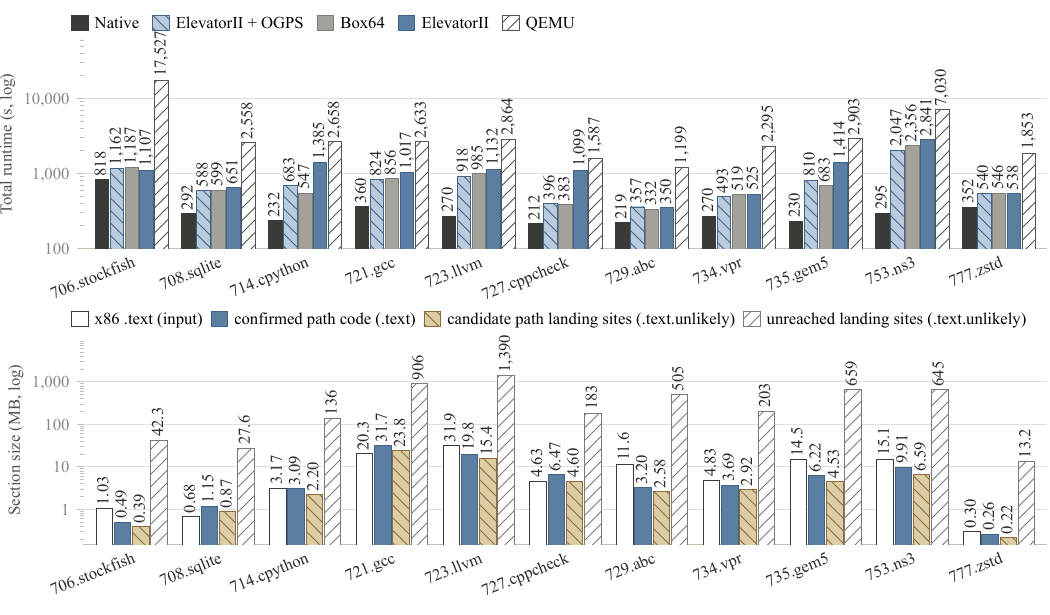}
    \caption{Runtime (top) and code sections of the \coolname{} + OGPS binary (bottom) per benchmark at \texttt{-O2}. The section split is the translator's own reachability classification.}
    \label{fig:runtime-layout-o2}
\end{figure*}
 
\section{Evaluation}
\label{sec:eval}
\textbf{Benchmarks.} We evaluate \coolname{} on SPEC CPU 2026~\cite{speccpu2026}, the current release of the SPEC CPU benchmarks, published in May 2026 as the successor of CPU 2017. We use the SPECrate 2026 Integer suite, 14 programs written in C and C++, and run each as a single copy on its reference inputs. The programs are real applications with large user bases, from compilers and interpreters to simulators and databases, selected by SPEC for diversity of application domains and code behavior~\cite{spec2026paper}. The 2026 integer suite also has larger code footprints and more complex control flow than its predecessor~\cite{spec2026comparison}. We omit the floating-point suite, whose backend-bound numeric programs would mainly require \coolname{} to support more instructions, engineering effort that tests no new part of the translator. \coolname{} currently does not support C++ exception handling, % (Section~\ref{sec:limitation}),
and three programs throw and catch exceptions during normal execution: 707.ntest while loading game records, 710.omnetpp to end a simulation run, and 750.sealcrypto while validating its parameters. We leave out these three and report the remaining 11.

\textbf{Baselines.} Our comparison includes what we call ``native binaries,'' which are the benchmarks compiled directly to AArch64 using GCC 13.3.0 with \texttt{-O2}. We compare these against the same sources compiled to x64 with GCC 13.3.0 at \texttt{-O2}, then processed by \coolname{} and two dynamic binary translators. Box64~\cite{box64} is a mature, actively developed user-space x64 emulator for AArch64 Linux. QEMU~\cite{qemu} in its user mode runs the x64 process, its libraries included, under its TCG translator. We use Box64 v0.4.1 in its dynamic-recompilation mode, its fast JIT configuration rather than its interpreter, and QEMU 8.2.2 as packaged with Ubuntu 24.04. Both translate while the program runs, whereas \coolname{} translates once and produces a standalone executable.

% \msof{LLVM XXXX with option \texttt{-o2}. We compare these against source code compiled to x86-64 using LLVM XXX with option XXX}

\textbf{Obfuscated Binaries.} We also evaluate \coolname{} on obfuscated binaries, whose control flow is deliberately hidden from static analysis using MBA and virtualization. We obfuscate the three C programs of the suite, 708.sqlite, 714.cpython, and 777.zstd, with Tigress~4.0.11~\cite{tigress}, a source-to-source obfuscator for C, and build them with the same flags as the original binaries.

\textbf{Experimental Setup.} All runs use a 64-core Ampere Altra (Arm Neoverse~N1) at 3.0~GHz with 64~GB of DDR4-3200, running Ubuntu~24.04.2 on Linux~6.8. Each core has 64~KB of L1 instruction cache, 64~KB of L1 data cache, and 1~MB of private L2. Every benchmark input runs three times in each execution mode, and we report the median wall-clock time. Hardware counters come from separate \texttt{perf~stat} runs with six Neoverse~N1 events each, so no event is multiplexed; their wall-clock times match the timing runs. OGPS profiles are collected on the same reference inputs. % (Section~\ref{sec:ogps}).
The evaluation answers four questions.
\begin{itemize}
  \item[\textbf{Q1}] How fast does \coolname{}-translated code run, with and without a profile, compared with Box64 and QEMU user mode, two widely used, actively developed open-source dynamic binary translators?
  \item[\textbf{Q2}] 
  %The static pipeline can only guess where a confirmed path should begin. A profile is the one source that shows where execution actually enters, so 
  How much does a single profiled run improve the translated binary, and through which mechanism?
  \item[\textbf{Q3}] %The same profile also shapes the binary's layout, beyond the runtime. 
  Superset translation emits code for every byte offset, While profiling-based reachability information separates the code observed as being executed from the rest. How large is each part?
  \item[\textbf{Q4}] Do these answers still hold when the input binary is obfuscated, its functions virtualized, its arithmetic encoded, and its direct branches routed through branch functions?
\end{itemize}

% Auto-generated by figures/runtime_table.py -- do not hand-edit
\begin{table}[b]
\centering
\small
\caption{Total runtime of the 11 SPEC CPU2026 benchmarks (45 reference inputs, \texttt{-O2}) per execution mode, and its ratio to native. Lower is better.}
\label{tab:runtime-total-2026}
\begin{tabular}{lrr}
\toprule
Execution mode & Total runtime (s) & Relative to native \\
\midrule
Native AArch64 & 3,550 & $1.00\times$ \\
\coolname{} + OGPS & 8,816 & $2.48\times$ \\
Box64 & 8,993 & $2.53\times$ \\
\coolname{} & 12,060 & $3.40\times$ \\
QEMU-user & 45,108 & $12.70\times$ \\
\bottomrule
\end{tabular}
\end{table}

\subsection{Results on SPEC CPU 2026}
\label{sec:runtime}

\label{sec:runtime-spec}
\textbf{Against the Dynamic Translators (Q1).} Over the 45 reference inputs, \coolname{} with OGPS finishes 2\% ahead of Box64 and 80\% ahead of QEMU (Table~\ref{tab:runtime-total-2026}). \coolname{} with OGPS runs between $1.4\times$ native (706.stockfish) and $6.9\times$ (753.ns3), Box64 between $1.5\times$ and $8.0\times$ on the same two benchmarks, and on 8 of the 11 the gap between the two translators is under 10\% (Figure~\ref{fig:runtime-layout-o2}). Both are slowest on the same five large programs. QEMU runs 3 to $15\times$ slower than \coolname{} with OGPS on every benchmark.

\begin{figure}[t]
    \centering
    \includegraphics[width=\columnwidth]{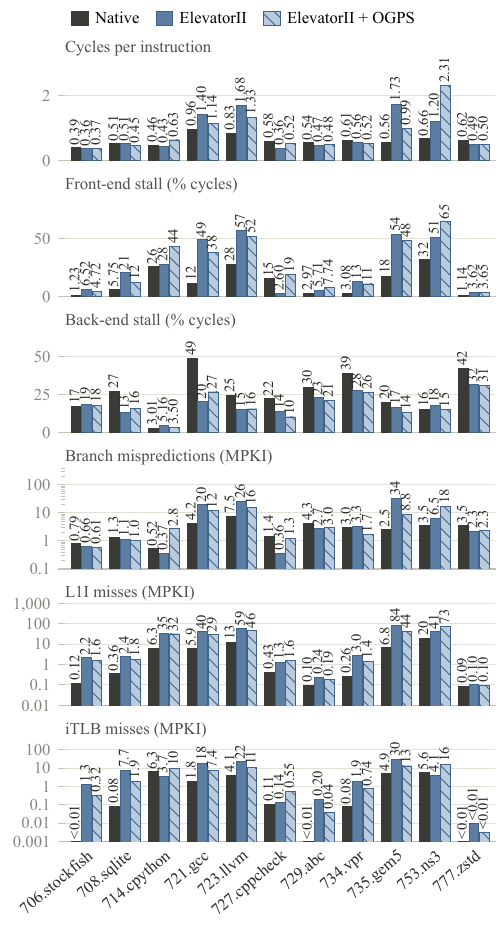}
    \caption{Hardware-counter metrics per benchmark at \texttt{-O2}. MPKI is misses per thousand instructions of the same run.}
    \label{fig:microarch-o2}
\end{figure}

Without the profile, \coolname{} trails Box64 by 34\%, and its slowdown over native ranges from $1.4\times$ to $9.6\times$. Two factors contribute to this range: how many instructions the translated program executes and how many cycles each one takes. On the first factor, most of the suite sits close together, 8 of the 11 benchmarks execute 1.5 to $2.3\times$ the native instructions. The three exceptions execute 5 to $8\times$ (727.cppcheck, 714.cpython, and 753.ns3), because much of the code they execute is entered through indirect control transfers and has no confirmed path, so it runs on the candidate path (Section~\ref{sec:candidate-path}). For these three, the profile observes many branch destinations the static pipeline does not know, 51.9\% and 51.1\% of the observed destinations in 727.cppcheck and 714.cpython and 36.9\% in 753.ns3 (Table~\ref{tab:ogps-heads-2026}). On the second factor, 7 of the 11 benchmarks keep a CPI at or below native, so their translated instructions run as efficiently as the compiler's own and their entire slowdown is the extra instructions. The four exceptions raise their CPI by 1.5 to $3\times$ (721.gcc, 723.llvm, 735.gem5, and 753.ns3), and Figure~\ref{fig:microarch-o2} separates the two causes. Their native runs already stress the instruction cache, and translation doubles the instructions and spreads them over the superset layout, so instruction-cache and TLB misses multiply. Every guest return and indirect branch is also dispatched through the address table instead of the hardware's return predictor, and mispredictions rise by 7 to $27\times$. Together these leave about half of their cycles stalled in the front end, against 12 to 32\% natively.

% Auto-generated by figures/pgo_heads_table.py -- do not hand-edit
\begin{table*}[t]
\centering
\small
\setlength{\tabcolsep}{5.5pt}
\caption{What the profile discovers per benchmark and what it buys. Discovered entries are observed branch destinations unknown to the static pipeline, each starting a confirmed path, counted out of all observed destinations and weighted by the share of sampled landings that enter through them. Instructions and CPI are relative to native, without $\rightarrow$ with the profile, and Speedup is the runtime reduction where the profile gives one.}
\label{tab:ogps-heads-2026}
\begin{tabular}{lrrrrrl}
\toprule
 & \multicolumn{2}{c}{Confirmed path entries the profile discovers} & & & & \\
\cmidrule(lr){2-3}
Benchmark & count/observed & share of landings & Instr. & CPI & Speedup & Gain \\
\midrule
706.stockfish & 22/1,328 (1.7\%) & 3.4\% & $1.5\rightarrow1.5$ & $0.93\rightarrow0.94$ & --- & --- \\
708.sqlite & 493/703 (70.1\%) & 96.1\% & $2.3\rightarrow2.3$ & $0.99\rightarrow0.88$ & 10\% & better code placement \\
714.cpython & 1,043/2,041 (51.1\%) & 77.4\% & $6.4\rightarrow2.2$ & $0.94\rightarrow1.37$ & 51\% & more code on confirmed paths \\
721.gcc & 1,807/36,546 (4.9\%) & 6.2\% & $1.9\rightarrow1.9$ & $1.46\rightarrow1.18$ & 19\% & better code placement \\
723.llvm & 1,084/25,641 (4.2\%) & 3.4\% & $2.1\rightarrow2.1$ & $2.02\rightarrow1.60$ & 19\% & better code placement \\
727.cppcheck & 3,570/6,885 (51.9\%) & 85.2\% & $8.2\rightarrow2.1$ & $0.63\rightarrow0.90$ & 64\% & more code on confirmed paths \\
729.abc & 2,083/3,166 (65.8\%) & 74.7\% & $1.8\rightarrow1.8$ & $0.88\rightarrow0.90$ & --- & --- \\
734.vpr & 91/4,069 (2.2\%) & 4.3\% & $2.1\rightarrow2.2$ & $0.91\rightarrow0.85$ & 6\% & better code placement \\
735.gem5 & 548/8,199 (6.7\%) & 4.4\% & $2.0\rightarrow2.0$ & $3.06\rightarrow1.75$ & 43\% & better code placement \\
753.ns3 & 5,418/14,671 (36.9\%) & 64.4\% & $5.3\rightarrow2.1$ & $1.80\rightarrow3.47$ & 28\% & more code on confirmed paths \\
777.zstd & 6/741 (0.8\%) & 0.1\% & $1.9\rightarrow1.9$ & $0.80\rightarrow0.80$ & --- & --- \\
\bottomrule
\end{tabular}
\end{table*}

\textbf{The Effect of OGPS (Q2).} OGPS improves 8 of the 11 benchmarks, and the Gain column of Table~\ref{tab:ogps-heads-2026} separates the two ways it helps. Each profile sample records one taken branch and the address it lands on. In the rows gaining more confirmed paths, a third to a half of the observed branch destinations were entries the static pipeline did not know. Each becomes a chain head, the code behind them moves from the candidate path onto confirmed paths, and the instruction column collapses from 5 to $8\times$ native to the suite's $2\times$ (727.cppcheck, 714.cpython, and 753.ns3). The runtime falls less than the instruction count, by 28 to 64\%, because the CPI column rises at the same time. 714.cpython also stands out in the mispredict row. Its mispredictions rise in every class of branch, direct branches included, so the predictor runs out of capacity on the denser confirmed path code rather than failing on any particular branch, and the cost stays near 5\% of its runtime. In the rows gaining fewer fetch misses, the instruction count does not move, whether the profile discovered many entries (70\% of 708.sqlite's observed destinations) or almost none (at most 7\% for 721.gcc, 723.llvm, 734.vpr, and 735.gem5). The same instructions run with a fifth to a half fewer instruction-cache misses and half to three quarters fewer instruction-TLB misses (Figure~\ref{fig:microarch-o2}, L1I and ITLB rows), and the runtime falls by 6 to 43\%. Branch mispredictions fall as well, by 38 to 74\% in every row but 708.sqlite's. The dispatch through the address table is unchanged, and the CPI still ends above native for 721.gcc, 723.llvm, and 735.gem5. The remaining three change by 5\% or less (706.stockfish, 729.abc, and 777.zstd), 706.stockfish for the worse with 4\% more instructions executed.

729.abc and 708.sqlite show the limit of discovery. Before the profile, a branch destination that no static disassembler can find runs on the candidate path. The profile observes these destinations, each becomes the entry of a confirmed path, and the instruction count should fall. %(Section~\ref{sec:ogps}).
However, when the candidate path at a discovered entry holds only a few instructions and is followed by a confirmed path, the conversion gains little. The chain built at the entry stops at the next branch target, %(Section~\ref{sec:confirmed-path}),
which caps the new confirmed path at those few instructions. A chain this short also leaves the optimizer nothing to combine, so its instructions run much as they did on the candidate path. Most of the discovered entries in both binaries are a single conditional branch followed immediately by a static entry. Their entries receive 75\% and 96\% of the sampled landings, yet the instruction counts do not change. Every candidate path instruction ends in a taken branch, so even a short candidate path appears often in the samples. 708.sqlite still gains 10\% through fewer fetch misses, and 729.abc has nothing left to gain, its instruction count already at $1.8\times$ native with a CPI below native.

\textbf{Binary Layout (Q3).}
\label{sec:layout}
\coolname{} owes its correctness and the size of the resulting binaries to the same decision, the translation of every byte offset. Elevator~\cite{elevator} demonstrates the cost for the same coverage, a translated \texttt{.text} about $54\times$ the natively compiled AArch64 one on this suite. \coolname{} emits at the same scale, 43 to $57\times$ per benchmark and $51\times$ over the suite. It accepts the cost rather than guess which decodes to drop. The code partitioning made possible by profiling enables us to give a concrete answer to how much noise the superset introduces.

\coolname{} follows direct control flow from the function entries and the observed landing sites, and marks every translation it reaches (Section~\ref{sec:ogps}). This reachability walk partitions the emitted code into three parts (Figure~\ref{fig:runtime-layout-o2}, bottom). Summed over the 11 binaries, confirmed path code is 1.8\% of the emitted code, and candidate path landing sites are another 1.3\%, among them the standalone copies of instructions the chains also contain. The remaining 96.9\% is code the walk never reaches, overlapping decodes of the same bytes and code nothing leads to. Per benchmark this share never falls below 93\%, from the 14~MB of 777.zstd to the 1.4~GB of 723.llvm.

The profile changes the layout in two ways. Without it, \coolname{} has no evidence of what runs, and every translation sits in the order of its input address, code that executes interleaved with overlapping decodes. With the profile, observed destinations become new entries with confirmed paths of their own, and an address never observed as a target does not need to sit near the code that runs. With the profile only the entries stay together, and everything else moves away (Algorithm~\ref{alg:ogps-layout}). Confirmed path code ends up packed into 1.8\% of the emitted bytes, and the fetch-miss gains of Q2 follow (Figure~\ref{fig:microarch-o2}, L1I and ITLB rows).

The partitioning rests on the profiling inputs, and a path the reference workloads never took may still run. This coverage limit is shared by any translator that relies on observed executions. \coolname{} therefore moves the never-reached code instead of dropping it, and such a path still finds its translation.

\textbf{Results on Obfuscated Binaries (Q4).}
\label{sec:runtime-obf}
In each program we obfuscate the hot functions of one source file with three Tigress transformations applied together. \textit{Virtualize} compiles a function into bytecode for a made-up instruction set and replaces its body with an interpreter that dispatches the bytecode through a switch. \textit{EncodeArithmetic} rewrites its arithmetic as mixed Boolean-arithmetic expressions. \textit{AntiBranchAnalysis} replaces each direct branch with a call to a branch function that overwrites its own return address with the branch target and returns. In the compiled binary the switch becomes a jump table and every branch function ends in a return to an address written at run time, so the obfuscated program takes indirect branches where the original took direct ones. This is the case the grid exists for. No static analysis can recover these targets, and \coolname{} needs none, because every byte offset already has a grid element to land on.

\begin{figure}[h]
    \centering
    \includegraphics[width=\columnwidth]{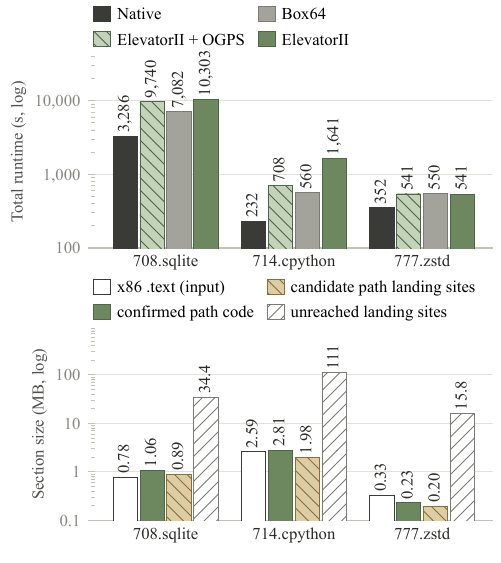}
    \caption{Runtime and code sections for the obfuscated programs.}
    \label{fig:runtime-layout-obf-o2}
\end{figure}

\coolname{} translates the obfuscated binaries with the pipeline of Figure~\ref{fig:overview} unchanged, Stage~2 included, and every translated binary produces the same output as its unobfuscated build. Figure~\ref{fig:runtime-layout-obf-o2} shows the runtimes. The slowdown follows the time each program spends in its obfuscated functions. In 708.sqlite they are the hot code: the obfuscated program runs $11\times$ longer than the unobfuscated one even natively, $12\times$ under Box64, and $16\times$ under \coolname{}; the profile recovers 5\%. 714.cpython spends little time there: native and Box64 stay within 3\% of their unobfuscated runtimes, \coolname{} within 18\% without the profile and 4\% with it. 777.zstd stays within 1\% in every configuration.

The gap on 708.sqlite is the cost of dispatch. The interpreter inside each obfuscated function reaches every bytecode operation through its jump table, one indirect jump per operation, and \coolname{} dispatches each of these jumps through the address table, as it does every indirect branch. The Stage~2 profile starts confirmed paths at the jump table's landing sites, but execution still enters each operation through the dispatch, which no profile can remove. The translated interpreter therefore runs a short confirmed path and pays a bounds check, a table load, and an indirect branch once per bytecode operation.

% The answers of the unobfuscated suite carry over. Box64 stays ahead where obfuscated code dominates, $2.2\times$ native against $3.0\times$ on 708.sqlite and $2.4\times$ against $3.1\times$ on 714.cpython, and the two are within 2\% on 777.zstd (Q1). The profile finds 480 new entry points in 708.sqlite, 173 in 714.cpython, and 139 in 777.zstd, receiving 81\%, 9\%, and 3\% of the sampled landings (Q2). Between 95 and 97\% of the emitted code is never reached (Q3, Figure~\ref{fig:runtime-layout-obf-o2}, bottom).

The answers of the unobfuscated suite carry over. Box64 stays ahead where the obfuscated functions dominate, $2.2\times$ the obfuscated native against \coolname{}'s $3.0\times$ on 708.sqlite and $2.4\times$ against $3.1\times$ on 714.cpython, and the two run within 2\% of each other on 777.zstd (Q1). The profile still finds what the static pipeline cannot: 480 new entries in 708.sqlite that receive 81\% of its sampled landings, 173 entries with 9\% in 714.cpython, and 139 with 3\% in 777.zstd, in the same order as the time the three programs spend in their obfuscated functions (Q2). The never-reached share of the emitted code stays between 95 and 97\% (Q3, Figure~\ref{fig:runtime-layout-obf-o2}, bottom).

The native AArch64 build is a reference, not a comparison: Tigress randomizes every run and emits architecture-specific branch functions, so the AArch64 build may differ from the x64 one.

\section{Related Work}

% \textbf{Disassembly and binary rewriting.} Rewriting a binary without its source requires knowing where its instructions begin, and Horspool and Marovac~\cite{horspoolAndMarovac1980detranslation} showed that deciding this in general is equivalent to the halting problem~\cite{turing1967halting}. Rewriters and lifters therefore lean on relocations, symbols, or an external disassembler~\cite{WilliamsKing2020,dinesh2020retrowrite,Yadavalli2019,Dinaburg2014}, on data-flow analyses~\cite{difederico2017revng}, or on probabilities~\cite{miller2019probabilistic}, and their heuristics fail on data in code and jump tables~\cite{pang2021sok} and on binaries built with unusual compiler flags~\cite{Cornucopia}. E9Patch~\cite{duck2020binary} patches in place without moving instructions. BinRec~\cite{altinay2020binrec} and WYTIWYG~\cite{parzefall2024you} lift only the code a traced run exercised, and Polynima~\cite{deshpande2024polynima} stops to relift at any indirect target it has not seen. Superset disassembly~\cite{Bauman2018SupersetDS} decodes at every byte offset and keeps every decode, so the rewritten program can be entered at any original address.

\textbf{Disassembly and Binary Rewriting.} Rewriting a binary without its source requires knowing where its instructions begin, which is generally undecidable for most architectures~\cite{horspoolAndMarovac1980detranslation}. Rewriters and lifters therefore lean on relocations, symbols, or an external disassembler~\cite{WilliamsKing2020,dinesh2020retrowrite,Yadavalli2019,Dinaburg2014}, on data-flow analyses~\cite{difederico2017revng}, or on probabilities~\cite{miller2019probabilistic}, and their heuristics don't always succeed on data in code and jump tables~\cite{pang2021sok} and on binaries built with unusual compiler flags~\cite{Cornucopia}. E9Patch~\cite{duck2020binary} patches in place without moving instructions, and BinRec~\cite{altinay2020binrec} lifts only the code a traced run exercised. Superset disassembly~\cite{Bauman2018SupersetDS} decodes at every byte offset and keeps every decode, so the rewritten program can be entered at any original address.

\textbf{Cross-ISA Translation.} LLBT~\cite{translator-for-arm1} lifts ARM to LLVM IR statically, with an address table for indirect branches, Lasagne~\cite{rocha2022lasagne} translates x86-64 to AArch64 with a proven mapping of its memory ordering onto fences, and Biotite~\cite{biotite} lifts RISC-V and dispatches indirect jumps through a switch over every instruction address. Dynamic translators see every target as it executes. QEMU~\cite{qemu} translates each basic block when first reached. HQEMU~\cite{hong2012hqemu} recompiles its hot traces through LLVM, HERMES~\cite{HERMES} optimizes the emitted host code rather than the IR, Pico~\cite{multicores} scales it to multicore guests, its translation rules have been learned from binaries compiled from the same source~\cite{LearnedTranslationRules}, and Risotto~\cite{risotto} adds proven fence mappings. Instrew~\cite{engelke2021llvmdbt} lifts to LLVM IR directly, and MAMBO-X64~\cite{MAMBO-X64} runs AArch32 on AArch64 and keeps calls paired with returns so the hardware return predictor works. Of the deployed x86-64 translators for AArch64, Rosetta~2~\cite{apple-rosetta-2,prj_champ_rosetta2} translates a whole image before its first run using heuristics to separate code and data and falls back to dynamic translation when it encounters missed jump targets at runtime, and Prism~\cite{prism} and Box64~\cite{box64} translate at run time. Arancini~\cite{arancini2026} combines a static and a dynamic translator over one IR. Elevator~\cite{elevator} translates instructions statically using superset disassembly, but doesn't optimize the result; \coolname{} builds on its insights.

% Elevator~\cite{elevator} carried superset disassembly across ISAs, with one translation per byte offset and an address table from every offset to its translation; \coolname{} builds on that substrate.

\textbf{Profile-guided Optimization.} FX!32~\cite{fx32} brought profiles to binary translation in 1998, emulating x86 programs on Alpha and translating, between runs, the procedures the profile had reached, and Dynamo~\cite{dynamo} optimizes native code as it runs, picking hot traces with counters so cheap that finer profiling predicts no better~\cite{lessIsMore}. Compilers work on the IR with the source at hand. They take the profile from instrumentation~\cite{ballLarus1994profiling} or hardware samples~\cite{autoFDO}, use it for block placement~\cite{profile-guided-code-positioning} and inlining, and can carry it across modules with link-time optimization~\cite{thinLTO}, and Codestitcher~\cite{codestitcher} lays out basic blocks across procedures when it recompiles the program from bitcode. The same profiles can also be applied after the compiler is done. Function ordering from sampled profiles~\cite{hfsort} improves on Pettis and Hansen's algorithm for data-center binaries, BOLT~\cite{BOLT} rewrites the linked binary, Propeller~\cite{propeller-google} relinks it from basic-block sections, DeduBB~\cite{DeduBB} merges duplicate cold blocks across procedures on either, and OCOLOS~\cite{ocolos} relays out a process while it runs. Twig~\cite{twig} uses the profile to insert Branch Target Buffer (BTB) prefetches rather than to move code. 

% \coolname{} starts from the binary, like the post-link tools, but lifts it to IR, like a compiler. It uses its profile for what no static analysis can supply, the landing sites of indirect control transfers, and keeps the input compiler's block order.

\section{Discussion of Limitations}
\label{sec:limitation}
\coolname{} translates only the code that is present in the binary, so it does not support programs that generate or modify their code at run time, such as JIT compilers. It does not yet support C++ exceptions. Throwing an exception unwinds the stack through the unwind tables of the binary, and the translated binary does not present the x64 frames to the unwinder in the form it expects. We exclude three SPEC CPU 2026 benchmarks for this reason. % (Section~\ref{sec:eval}).

Furthermore, \coolname{} supports single-threaded programs only. x64 assumes a strongly ordered memory model and AArch64 a weakly ordered one, so a multi-threaded translation requires fences wherever the x64 ordering is relied upon, as previously explored in Lasagne, Risotto, and Arancini~\cite{rocha2022lasagne,risotto,arancini2026}. Instead of adopting their approach of inserting fences, an alternative direct way to overcome this limitation would be to use one of the emerging custom ARM chip implementations~\cite{tosting,corbet2024tso,apple-rosetta-tso} that support an optional or switchable Total Store Order (TSO) memory model variant that mimics x86 memory semantics.

At first glance, the enormous space overhead of our approach appears problematic, but as our experiments show, the working code set in our experiments is only a small fraction of the total generated code size. We plan to study further optimizations of the ``unlikely code'' ELF section layout towards minimizing memory paging traffic.

\section{Conclusion}

Executing existing binaries for one ISA on a different ISA has traditionally been achieved by a combination of emulation and just-in-time compilation. This typically requires a substantial software support infrastructure at runtime and encompasses dynamic optimization decisions that are hard to predict or reproduce.

We introduce a novel approach that first creates a static ``grid'' translation in which every instruction of the input program, including even partially overlapping instructions, is implemented by a corresponding basic block in the output program that faithfully models all flag computations and other side effects of the original instruction. These per-instruction translations form the candidate paths. We then add optimized confirmed paths, selected by observation-guided path selection (OGPS) on the target platform. Our resulting output binaries consistently perform as well or better than mature state of the art emulators with hybrid static/dynamic compilation engines. Our approach is also safer, since the resulting output binaries are generated entirely statically, require no additional run-time support, and can be exhaustively tested before deployment. 

\begin{acks}

This work is partially supported by ERC Grant REVITALIZE, Agreement No. 101199828. Funded by the European Union. Views and opinions expressed are however those of the authors only and do not necessarily reflect those of the European Union or the European Research Council Executive Agency. Neither the European Union nor the granting authority can be held responsible for them.

This material is based upon work partially supported by the Defense Advanced Research Projects Agency (DARPA) under Contract No.~140D04-23-C-0070. Any opinions, findings and conclusions or recommendations expressed in this material are those of the authors and do not necessarily reflect the views of the Defense Advanced Research Projects Agency (DARPA), its Contracting Agent, or any other agency of the U.S. Government.

\end{acks}

\bibliographystyle{ACM-Reference-Format}
\bibliography{ElevatorII-arXiv}

\end{document}